\documentclass[10pt]{article}
\usepackage{amsfonts, amssymb, amsthm,amsmath}
\usepackage{epsfig}
\usepackage{color}

\newtheorem{tw}{Theorem}
\newtheorem{de}{Definition}
\newtheorem{co}{Corollary}
\newtheorem{pr}{Proposition}

\newcommand{\be}{\begin{equation}}
\newcommand{\ee}{\end{equation}}
\newcommand{\bea}{\begin{eqnarray}}
\newcommand{\eea}{\end{eqnarray}}

\begin{document}


\begin{center}
{\LARGE{\bf{Abelian connection in Fedosov deformation quantization. \\
\vskip0.50cm
 I. The $2$-dimensional phase space }}}
\end{center}

\begin{center}
\vskip0.25cm

{\bf Jaromir  Tosiek}  \\

\vskip0.25cm
{\em  Institute of Physics}\\ {\em Technical University of Lodz}\\ {\em ul. Wolczanska 219, 93-005 Lodz}\\
{\em Poland}

\vskip0.15cm

e-mail:  tosiek@p.lodz.pl

\vskip0.25cm\centerline{\today}

\begin{abstract}
General properties of an Abelian connection in Fedosov deformation quantization are investigated. Definition and criterion of being  a finite formal  series 
for an Abelian connection are presented. A proof that in $2$-dimensional ($2$-D) case the Abelian connection is an ifinite formal series is done. 
\end{abstract}

\end{center}
{\bf Keywords}: Fedosov deformation quantization, Abelian connection.
\newline
{\bf PACS} numbers: 02.40.Hw, 03.65.Ca

\section{Introduction}
Deformation quantization on a phase space ${\mathbb R}^{2n}$ was invented in the middle of the previous century.  Moyal~\cite{MO49} using  results obtained  by Weyl~\cite{WY31}, Wigner~\cite{WI32} and Groenewold~\cite{GW46} presented quantum mechanics  perceived as  statistical theory. His paper contained not only general considerations but also explicit  formulas defining the $*$-product of observables and so called Moyal bracket being the counterpart of the commutator of operators.  

The first succesful generalization of Moyal results in case of phase space different from  ${\mathbb R}^{2n}$ appeared in 1977 when Bayen {\it at al} \cite{baf, bay} proposed an axiomatic version of deformation quantization. In those articles 
quantum mechanics gained a new aspect - as a deformed version of classical physics. 
Unfortunately, in contrast to Moyal, quoted  authors did not present universal `computable' model of their idea. Since that their results were  applicable only in some special cases like a harmonic oscillator.

One of  realisations of Bayen {\it at al} quantization programme is so called Fedosov deformation quantization \cite{6, 7}. The Fedosov construction is  algebraic and it can be applied easily to solve some problems like harmonic oscillator \cite{ja, my1} or $2$-D symplectic space with constant curvature tensor \cite{ my2}. Great advantage of that method is the fact that 
computations may be done by computer programmes. 

Since the publication of a proof that all $*$-products on a symplectic manifold are equivalent to the Fedosov product \cite{nest}, this quantization method can be treated just as an algorithm to do computations in deformation quantization at all.

In Fedosov quantization we work with formal series. There is no general method  to write these series in a compact form. 
Series of compact form appear inter alia when they contain finite number of terms.   
In that case $*$- product of functions can be calculated strictly.

Fedosov deformation quantization is based on two recurrent equations, which  generate  formal series. The first one  is the formula defining an Abelian connection, the second - relation introducing a series representing an observable. In this paper we deal with  the Abelian connection.

Because the answer, when the Abelian connection  is a finite series, depends on the dimension of the phase space of a system, we divide our considerations in two papers. In this one we present the necessary and sufficient condition for the Abelian connection to be a finite series. We prove that in $2$-D  case the Abelian connection cannot be represented by such a  series. In the second part  we will deal with more dimensional phase spaces.
 
 Our considerations are devoted only to Abelian connections determined by the iteration process proposed by Fedosov. Another kind of Abelian connection on K\"{a}hler  symmetric manifolds can be found in \cite{tama}.

In all of formulas, in which summation limits are obvious, we use Einstein summation convention.

\section{Foundations of  Fedosov deformation quantization}

\setcounter{pr}{0}
\setcounter{co}{0}
\setcounter{tw}{0}
\setcounter{de}{0}
\setcounter{equation}{0}

All  facts presented in this section have been published in some books or papers. 
We set them together to simplify following our considerations. 
 Moreover, we modify   notation a bit comparing to the original Fedosov article.

The starting point of the deformation quantization according to the Fedosov rules is a symplectic manifold equipped with some connection known as `symplectic'. 
 Reader interested in details of symplectic geometry is pleased to look them up in \cite{7}, \cite{my1},
\cite{va}, \cite{gel}.

Let $({\cal M}, \omega)$ be a $2n$-D symplectic manifold and ${\cal A}= \{({\cal U}_z,\phi_z)\}_{z \in J}$  an atlas on ${\cal M}.$ By $\omega$ we mean the symplectic $2$-form. Since we work only with symplectic manifolds, in our paper we will denote a manifold $({\cal M}, \omega)$ just by ${\cal M}.$
\begin{de}
{\bf The symplectic connection} $\Gamma$ on ${\cal M}$ is a torsion free connection  locally  satisfying  conditions
\be
\label{d00}
\omega_{ij;k}=0,\;\;\; 1 \leq i,j,k \leq 2n, 
\ee
where a semicolon `$\: ;$' stands for the covariant derivative.
\end{de}
In any Darboux coordinates the system of equations (\ref{d00}) reduces to
\be
\label{d1.1}
\omega_{ij;k}= - \Gamma^l_{ik}\omega_{lj} - \Gamma^l_{jk}\omega_{il}= -\Gamma_{jik}+ \Gamma_{ijk}=0,
\ee
where
$\Gamma_{ijk} \stackrel{\rm def.}{= } \Gamma^l_{jk}\omega_{li}.
$

As it can be seen from (\ref{d1.1}), coefficients $\Gamma_{ijk}$ are symmetric with respect to indices $\{i,j,k\}.$ The number of independent elements $\Gamma_{ijk}$ equals
$ \left(
\begin{array}{c}
2n+2\\2n-1
\end{array}
\right).
$ A symplectic connection exists on any symplectic manifold.

\begin{de}
The symplectic manifold ${\cal M}$ equipped with the symplectic connection $\Gamma$
 is called the {\bf  Fedosov manifold} $({\cal M}, \Gamma).$
\end{de}

In  Darboux coordinates the symplectic curvature tensor $R_{\Gamma}$
\[
(R_{\Gamma})_{ijkl}=\frac{\partial \Gamma_{ilj}}{\partial q^k} - \frac{\partial \Gamma_{ijk}}{\partial q^l } + \omega^{mp} \Gamma_{plj}\Gamma_{ikm}-  \omega^{mp} \Gamma_{pjk}\Gamma_{ilm}.
\]
Between the tensor $\omega^{ij}$ and the symplectic form $\omega_{jk}$ the relation holds
$
\omega^{ij} \omega_{jk}= \delta^i_k.
$

 The symplectic curvature tensor $R_{\Gamma}$ is symmetric in two first indices $
(R_{\Gamma})_{ijkl}=(R_{\Gamma})_{jikl}$ and 
antisymmetric  in two last indices 
$
(R_{\Gamma})_{ijkl}=- (R_{\Gamma})_{ijlk}.
$
On a $2n$-D Fedosov manifold $({\cal M}, \Gamma)$ a number of independent components of the tensor $R_{\Gamma}$ equals
$\frac{1}{2}n(n+1)(2n+1)(2n-1).$

\hspace{0.25cm}

Let $\hbar$ denote some positive parameter and $X^1_{\tt p}, \ldots, X^{2n}_{\tt p}$ components of an arbitrary vector ${\bf X}_{\tt p}$ belonging to the tangent space $T_{\tt p} {\cal M}$
to the symplectic manifold ${\cal M}$ at the point ${\tt p}. $ Components $X^1_{\tt p}, \ldots, X^{2n}_{\tt p}$ are written in the natural basis $\left( \frac{\partial}{\partial q^i}\right)_{\tt p}$ determined by the chart $({\cal U}_z,\phi_z)$ such that ${\tt p} \in {\cal U}_z.$

In the point ${\tt p}$ we introduce a formal series 
\be
\label{d3}
a \stackrel{\rm def.}{=}\sum_{l=0}^{\infty}\hbar^k a_{k, i_1 \ldots i_l} X^{i_1}_{\tt p} \ldots X^{i_l}_{\tt p}, \;\; k \geq 0.
\ee
For $l=0$ we put $a = \hbar^k a_k.$ By $a_{k, i_1 \ldots i_l}$ we denote components of a covariant tensor symmetric with respect to indices $\{i_1 \ldots, i_{l}\}$ taken in a basis $dq^{i_1} \odot \ldots \odot dq^{i_l}. $

A part of the series $a$ standing at $\hbar^k$ and containing $l$ components of the vector ${\bf X}_{\tt p}$ will be denoted by
$a[k,l]$ so 
$
a= \sum_{k=0}^{\infty}\sum_{l=0}^{\infty}\hbar^k a[k,l].
$
 The {\bf degree} $\deg(a[k,l]) $ of the component $a[k,l]$ is the sum $2k+l.$
 The degree of the series $a$ is the maximal degree of its  nonzero components $a[k,l].$

Let $P^*_{\tt p}{\cal M}[[\hbar]]$ be a set of all elements $a$ of the kind (\ref{d3}) at the point ${\tt p}. $ The set 
$P^*_{\tt p}{\cal M}[[\hbar]]$ is a linear space over ${\mathbb C}.$
\begin{de}   
The product $\circ: P^*_{\tt p}{\cal M}[[\hbar]] \times P^*_{\tt p}{\cal M}[[\hbar]]\rightarrow P^*_{\tt p}{\cal M}[[\hbar]]$ of two elements $
a, b \in P^*_{\tt p}{\cal M}[[\hbar]]$ is the mapping
\be
\label{6}
a \circ b
\stackrel{\rm def.}{=}  \sum_{t=0}^{\infty} \frac{1}{t!}\left(\frac{i\hbar}{2}\right)^t\omega^{i_1 j_1} \cdots \omega^{i_t j_t} \:\frac{\partial^{ t}  a }{\partial X^{i_1}_{\tt p}\ldots\partial X^{i_t}_{\tt p} }\:\frac{\partial^{t}  b }{\partial X^{j_1}_{\tt p}\ldots\partial X^{j_t}_{\tt p} }.
\ee
\end{de}
The pair $(P^*_{\tt p}{\cal M}[[\hbar]],\circ) $ is a noncommutative associative algebra called the {\bf Weyl algebra}. The $\circ$-product 
does not depend on the chart and is in general nonabelian. Moreover, 
 for all \\ $a, b \: \in \:(P^*_{\tt p}{\cal M}[[\hbar]],\circ)$ the relation holds
\[
 \deg(a \circ b)= \deg(a) + \deg(b).
\]

\begin{de}
\label{tos0}
 A {\bf Weyl bundle} is a triplet $({\cal P^*M}[[\hbar]],\pi, {\cal M}),$ where 
\[
{\cal P^*M}[[\hbar]] \stackrel{\rm def.}{=} \bigcup_{{\tt p} \in {\cal M}}   (P^*_{\tt p}{\cal M}[[\hbar]],\circ)
\]
is a differentiable manifold called a total space, ${\cal M}$ is a base space  and   $\pi:{\cal P^*M}[[\hbar]] \rightarrow {\cal M}$  a projection.
\end{de}
The Weyl bundle is a vector bundle in which the typical  fibre is also an algebra.
\begin{de}
{\bf The $n$-differential form} with value in the Weyl bundle is the form written locally 
\be
\label{e1}
a= \sum_{l=0}^{\infty}\hbar^k a_{k, i_1 \ldots i_l, j_1 \ldots j_m}(q^1, \ldots, q^{2n}) X^{i_1} \ldots  X^{i_l}dq^{j_1} \wedge
\cdots \wedge dq^{j_m},
\ee
where $0 \leq m \leq 2n.$ Now
$a_{k, i_1 \ldots i_l, j_1 \ldots j_m}(q^1, \ldots, q^{2n}) $ are components of smooth tensor fields on ${\cal M}$ and \\
$ C^{\infty}({\cal TM})\ni {\bf X}\stackrel{\rm locally }{=} X^i \frac{\partial}{\partial q^i} $ is a smooth vector field on ${\cal M}.$
\end{de}
 For simplicity we will omit variables $(q^1, \ldots, q^{2n}).$ 

Let $\Lambda^m$ be a smooth field of $m$-forms on the symplectic manifold ${\cal M}.$ Forms of the kind (\ref{e1}) are smooth sections of the direct sum
$ {\cal P^*M}[[\hbar]] \otimes \Lambda \stackrel{\rm def.}{=} \oplus_{m=0}^{2n}({\cal P^*M}[[\hbar]] \otimes \Lambda^m)
$.

The {\bf projection} $\sigma(a)$ of $a \in C^{\infty}({\cal P^*M}[[\hbar]] \otimes \Lambda^0)$ means $a|_{{\bf X}= {\bf 0}}$.

\begin{de}
{\bf The commutator} of forms $a \in C^{\infty}({\cal P^*M}[[\hbar]] \otimes \Lambda^{m_1})$ and  $b \in C^{\infty}({\cal P^*M}[[\hbar]] \otimes \Lambda^{m_2})$ is a form $[a,b] \in C^{\infty}({\cal P^*M}[[\hbar]] \otimes \Lambda^{m_1+m_2})$ defined by
\be
\label{e3}
[a,b] \stackrel{\rm def.}{=} a \circ b - (-1)^{m_1 \cdot m_2}b \circ a.
\ee
\end{de}

A form $a \in C^{\infty}({\cal P^*M}[[\hbar]] \otimes \Lambda )$ is called {\bf central}, if for every  $b \in C^{\infty}({\cal P^*M}[[\hbar]] \otimes \Lambda)$ the commutator $[a,b]$ vanishes.
Only forms not containing $X^i$'s are central in the Weyl algebra.

\begin{de}
The antiderivation operator 
$
 \delta:C^{\infty}({\cal P^*M}[[\hbar]] \otimes \Lambda^{m}) \rightarrow  C^{\infty}({\cal P^*M}[[\hbar]] \otimes \Lambda^{m+1})
$
 is defined by
\be
\label{e4}
\delta a \stackrel{\rm def.}{=}dq^k \wedge \frac{\partial a}{\partial X^k}.
\ee
\end{de}
The operator $\delta$ lowers the degree $\deg (a)$  of the  elements of  
${\cal P^*M}[[\hbar]]  \Lambda$ by $1$.

Every two forms $a \in C^{\infty}({\cal P^*M}[[\hbar]] \otimes \Lambda^{m_1})$ and $b \in C^{\infty}({\cal P^*M}[[\hbar]] \otimes \Lambda)$ satisfy
\be
\label{e5}
\delta(a \circ b)= (\delta a)\circ b + (-1)^{m_1} a \circ (\delta b).
\ee

\begin{de} 
The operator $\delta^{-1}:C^{\infty}({\cal P^*M}[[\hbar]] \otimes \Lambda^m) \rightarrow C^{\infty}({\cal
P^*M}[[\hbar]]
\otimes \Lambda^{m-1})$ is defined by
\be
\label{delta2}
\delta^{-1} a = \left\{ \begin{array}{ccl}
&\frac{1}{l+m}\: X^k \frac{\partial }{\partial q^k}\rfloor a \qquad  &{\rm for} \;\;\; l+m>0,   \\[0.35cm]
 & 0 \qquad &{\rm for} \;\;\; l+m=0,
\end{array}\right.
\ee
where $l$ is  the degree of $a$ in $X^i$'s i.e. the number of $X^i$'s. 
\end{de}
$\delta^{-1}$ raises the degree of the forms of ${\cal P^*M}[[\hbar]] \Lambda$ in the Weyl algebra by 
$1$.

The linear operators   $\delta$ and $\delta^{-1}$ do not depend on
the choice of local coordinates and have the following properties:

(i)
$
\delta^2 = (\delta^{-1})^2=0;
$

(ii) let us assume that indices $i_1, \ldots ,i_{l}$ and $j_1, \ldots, j_m$ are arbitrary but fixed. For the monomial

\hspace{0.5cm}
 $ X^{i_1}\ldots X^{i_l} dq^{j_1} \wedge \ldots \wedge dq^{j_m}$ we have
\[
(\delta \delta^{-1} + \delta^{-1} \delta)X^{i_1}\ldots X^{i_l} dq^{j_1}
\wedge \ldots \wedge dq^{j_m} = X^{i_1}\ldots X^{i_l}
dq^{j_1} \wedge \ldots \wedge dq^{j_m}.
\]

The straightforward consequence of the linearity and the decomposition of monomials is the Hodge decomposition of the form $a \in C^{\infty}({\cal P^*M}[[\hbar]]\otimes \Lambda)$ as shows the next theorem.
\begin{tw}
\cite{6}, \cite{7}
For every $a \in C^{\infty}({\cal P^*M}[[\hbar]]\otimes \Lambda)$ 
\be
\label{dec1}
a= \delta \delta^{-1}a + \delta^{-1} \delta a +a_{00},
\ee
where $ a_{00} $ is a smooth function on the symplectic manifold ${\cal M}$.
\end{tw}

\begin{de}
{\bf The exterior covariant derivative } $\partial_{\gamma}$ of the form $ a \in C^{\infty}({\cal P^*M}[[\hbar]] \otimes \Lambda^m )$ determined by a connection $1$-form 
$\gamma \in C^{\infty}({\cal P^*M}[[\hbar]] \otimes \Lambda^1)$ is the linear operator \\
$
\partial_{\gamma} : C^{\infty}({\cal P^*M}[[\hbar]] \otimes \Lambda^m ) \rightarrow C^{\infty}({\cal P^*M}[[\hbar]] \otimes \Lambda^{m+1} ) 
$
defined in a Darboux chart by the formula
\be
\label{f1}
\partial_{\gamma} a \stackrel{\rm def.}{=}da + \frac{1}{i \hbar}[\gamma,a].
\ee
\end{de}
In case when $\gamma$ represents the symplectic connection, we put
\be
\label{f1.1}
\gamma \stackrel{\rm denoted}{=}\Gamma= \frac{1}{2}\Gamma_{ijk}X^iX^j dq^k.
\ee

 The curvature form $R_{\gamma}$ of a connection $1$-form $\gamma$
in a Darboux chart can be expressed by the formula
\be
\label{f4}
R_{\gamma}= d \gamma + \frac{1}{2 i \hbar}[\gamma, \gamma]=  d \gamma + \frac{1}{ i \hbar}\gamma \circ  \gamma. 
\ee

Hence the second covariant derivative
$ \partial_{\gamma}(\partial_{\gamma}a)= \frac{1}{i \hbar}[R_{\gamma},a]$.

The crucial role in the Fedosov deformation quantization is played by an  Abelian connection $\tilde{\Gamma}.$ From the definition by the {\bf Abelian} connection we mean such connection $\tilde{\Gamma}$ whose  curvature form $R_{\tilde{\Gamma}}$ is  central  so
$
\partial_{\tilde{\Gamma}}(\partial_{\tilde{\Gamma}}a)=0
$
for every $a \in C^{\infty}({\cal P^*M}[[\hbar]] \otimes \Lambda).$

The Abelian connection proposed by Fedosov is of the form
\be
\label{f6}
\tilde{\Gamma}= \omega_{ij}X^i dq^j + \Gamma + r.
\ee
Its curvature 
\be
\label{f7}
R_{\tilde{\Gamma}}= - \frac{1}{2} \omega_{j_1 j_2}dq^{j_1} \wedge dq^{j_2} + R_{\Gamma}- \delta r + \partial_{\Gamma}r + \frac{1}{i \hbar}r \circ r.
\ee
The requirement that the central curvature $2$-form 
$
R_{\tilde{\Gamma}}= - \frac{1}{2} \omega_{j_1 j_2}dq^{j_1} \wedge dq^{j_2}
$
 means that we look for the solution of the equation
\be
\label{f8}
  \delta r=R_{\Gamma} + \partial_{\Gamma}r + \frac{1}{i \hbar}r \circ r.
\ee
Fedosov proved (see \cite{6}, \cite{7}) the following theorem.
\begin{tw}
\label{no0}
The equation (\ref{f8}) has a unique solution
\be
\label{2}
r = \delta^{-1} R_{\Gamma} + \delta^{-1} \left( \partial_{\Gamma} r + \frac{1}{i \hbar}r \circ r \right)
\ee
fulfilling the following conditions
\be
\label{h1}
\delta^{-1}r=0\;\;\; , \;\;\; \deg (r) \geq 3.
\ee
 \end{tw}

We work only with the Abelian connection of the form (\ref{f6}) with the correction $r$ defined by (\ref{2}) and fulfilling  (\ref{h1}). 
The general solution of (\ref{f8}) was published in \cite{ja3}.
The Abelian connection on K\"{a}hler symmetric manifolds proposed by Tamarkin \cite{tama} does not fulfill the condition $\delta^{-1}r=0.$

\begin{de}
The subalgebra ${\cal P^*M}[[\hbar]]_{\tilde{\Gamma}}  \subset C^{\infty}( {\cal P^*M}[[\hbar]] \otimes \Lambda^0 )$ consists of flat sections i.e. such that
$
\partial_{\tilde{\Gamma}}a=0.
$
\end{de}
\begin{tw}
\cite{6}, \cite{7}
For any $a_0 \in C^{\infty}(\cal{M}) $ there exists a unique section $a \in {\cal P^*M}[[\hbar]]_{\tilde{\Gamma}} $ such that the projection  
$ \sigma(a)=a_0.$ 
\end{tw}
Applying the operator $\delta^{-1}$ it follows from the Hodge decomposition (\ref{dec1}) that
\be
\label{nowy09}
a= a_0 + \delta^{-1} \left( \partial_{\Gamma}a + \frac{1}{i \hbar}[r,a]\right).
\ee

Using the one-to-one correspondence between ${\cal P^*M}[[\hbar]]_{\tilde{\Gamma}}$ and $C^{\infty}(\cal{M})$ we introduce an associative star product `$*$' of functions $a_0,b_0 \in C^{\infty}(\cal{M})$ 
\be
\label{jeszcze}
a_0 * b_0 \stackrel{\rm def.}{=} \sigma( \sigma^{-1}(a_0) \circ  \sigma^{-1}(b_0)).
\ee
The $*$- product (\ref{jeszcze}) fulfills axioms of the star product in deformation quantization  and is interpreted as the quantum multiplication of observables. 


\section{Properties of the Abelian connection }

\setcounter{pr}{0}
\setcounter{co}{0}
\setcounter{tw}{0}
\setcounter{de}{0}
\setcounter{equation}{0}

In this section  we present  general features of the Abelian connection constructed according to Fedosov procedure and consider   conditions under which the correction $r$ is a finite formal series. 

\subsection{Connections in the Weyl bundle }
 Let ${\cal P^*M}[[\hbar]]$ be the Weyl algebra bundle  equipped with some connection determined by $1$-form $\gamma$. We do not assume that $\gamma$ represents any Abelian or symplectic connection.
 
\begin{pr}
\label{g2}
Every  connection $\gamma \in C^{\infty}({\cal P^*M}[[\hbar]] \otimes \Lambda^1 )$ such that $\delta \gamma=0$  satisfies $\delta R_{\gamma} =0.$
\end{pr}

\underline{Proof}
\newline
For every $a \in C^{\infty}({\cal P^*M}[[\hbar]] \otimes \Lambda) $
\be
\label{g1}
d(\delta a)+ \delta(da)=0.
\ee
Hence the condition $\delta \gamma=0$  and  the property (\ref{g1}) 
 gives
\be
\label{g1..1}
\delta( d\gamma)=0.
\ee
Using  formula (\ref{f4}) describing the curvature form
 we obtain
\[
\delta R_{\gamma} =
\delta(d\gamma + \frac{1}{i \hbar} \gamma \circ \gamma)\stackrel{\rm (\ref{e5})}{=} \delta(d\gamma) + \frac{1}{i \hbar} 
\left(\delta(\gamma)\circ \gamma - \gamma \circ \delta(\gamma)\right).
\]
From the assumption $\delta \gamma=0$ and  equation (\ref{g1..1}) 
we see that indeed $\delta R_{\gamma}=0.$
 \rule{2mm}{2mm}

The straightforward consequence of proposition {\bf \ref{g2}} and decomposition (\ref{dec1}) is the following corollary.
\begin{co}
\label{co1}
If the connection form $\gamma$ fulfills the condition $\delta \gamma =0$ then its curvature $2$- form \\
$
R_{\gamma}= \delta \delta^{-1}R_{\gamma}.
$
\end{co}
Hence,
for  a connection $\gamma$ such that $\delta \gamma =0$ we have  $R_{\gamma} = 0$ if and only if 
$\delta^{-1}R_{\gamma} = 0.$

Let us use the above corollary to the symplectic connection represented by the $1$-form $\Gamma$  ( see (\ref{f1.1})).
Since the fact that coefficients $\Gamma_{ijk}$ are symmetric in indices $\{i,j,k\},$ we obtain that
$
\delta \Gamma=0.
$

Applying  corollary {\bf \ref{co1}} we conclude that
\begin{pr}
\label{tw1}
Two symplectic curvature forms $R_{\Gamma 1}$ and $R_{\Gamma'}$ defined by symplectic connections  $\Gamma $ and $\Gamma'$ respectively, are equal if and only if 
$\delta^{-1}R_{\Gamma }= \delta^{-1}R_{\Gamma'}.$ 
\end{pr}
From proposition {\bf \ref{tw1}} we see that geometry of a symplectic space can be characterized by a tensor $(R_{\Gamma})_{ijkl}$ symmetric in indices $\{i,j\}$ and antisymmetric in $\{k,l\}$ or, equivalently, by a tensor $(\delta^{-1}R_{\Gamma})_{ijkl}$ symmetric in indices $\{i,j,k\}.$

\vspace{0.5cm}
Let us consider  the structure of  equation (\ref{2}).
Its solution fulfilling  conditions (\ref{h1})  can be found by iteration method \cite{6,7}
\be
\label{p8}
r_{\bf 0}\stackrel{\rm def.}{=}0, \;\;
r_{\bf s}= \delta^{-1}\left(R_{\Gamma}+ \partial_{\Gamma}r_{\bf s-1}+ \frac{1}{i \hbar} r_{\bf s-1} \circ r_{\bf s-1}\right),\;\;{\bf s}=1,2, \ldots,  
\ee
The  component  of  $r$ of the lowest degree is $\delta^{-1}R_{\Gamma}$, $\deg(\delta^{-1}R_{\Gamma})=3$ and it is  the only one term of that degree. Hence, 
the solution  of (\ref{2}) can be written in a form  
\be
\label{z2}
r= \delta^{-1}R_{\Gamma} + \sum_{z=4}^{\infty}\sum_{k=0}^{[\frac{z}{2}]}\hbar^{k} r_m[k, z- 2k]dq^m.
\ee
By $[\frac{z}{2}]$ we denote the maximal integral  number not bigger than $\frac{z}{2}$. The symbol $r_m[k, z- 2k]dq^m$ means a $1$-form containing $(z-2k)$ $X^i$'s and standing at $\hbar^k.$ 

From proposition {\bf \ref{tw1}} we conclude that  if $R_{\Gamma } \neq R_{\Gamma'}$ then corrections $r$ determined by connections $\Gamma $ and $\Gamma'$ respectively are  different. \label{aa1}  We deduce that $z-2k \geq 1,$ because each component of $r_{\bf s}$ contains one or more $X$'s. Moreover, the product $r_{\bf s-1} \circ r_{\bf s-1}$ generates only odd powers of $\hbar$ so the index $k$ in (\ref{z2}) is even.

Therefore,  formula (\ref{z2}) can be written in the following form
\be
\label{z2.2}
r= \delta^{-1}R_{\Gamma} + \sum_{z=4}^{\infty}\sum_{k=0}^{[\frac{z-1}{4}]}\hbar^{2k} \:r_m[2k, z- 4k]dq^m.
\ee

 In case when $\deg(r)= d, \;\; d \in {\cal N} $ we say that $r$ is a {\bf finite formal series}.
For $\deg(r)= \infty$ we deal with an infinite series.

If in an arbitrary chart   
some terms $r_m[2k, z- 4k]dq^m$ for fixed $k$ and $z$ do not disappear, the same happens in any other chart. This statement follows from the fact that the $1$- form $r_m[2k, z- 4k]dq^m$ is determined by  tensor components $r_{i_1 \ldots i_{z-4k},m}\:,\;\; 1 \leq i_1, \ldots,i_{z-4k},m \leq \dim{\cal M}.$ Moreover, from the same reason 
\begin{co} 
At an arbitrary  point ${\tt p} \in {\cal M}$ the fact that the series $r$ is finite  does not depend on a chart.
\end{co}
\begin{co}
At an arbitrary point ${\tt p} \in {\cal M}$ inequalities
$ \;\partial_{\Gamma} r_m[l,u]dq^m \neq 0 , \; r_m[l,u]dq^m \circ r_j[p,k]dq^j  \neq 0 
$ 
are true in each chart.
\end{co}

\subsection{Finite Abelian connection}
\label{fac}

Now we consider in which cases an Abelian connection is represented by a finite formal series. This situation is eligible because  then Fedosov method may lead to strict quantization of a system. By $r[z]$ we will denote the  component $r[z]\stackrel{\rm def.}{=}\sum_{k=0}^{[\frac{z-1}{4}]}\hbar^{2k}r_{m}[2k,z-4k]dq^m, \;\;z \geq 3$   of $r$   of the degree $z.$

 As it was proved \cite{olga}, \cite{vais2}
\[
r[3]= \delta^{-1}R_{\Gamma},
\]
\be
\label{no1}
r[z]=\delta^{-1}\left( \partial_{\Gamma}r[z-1]+ \frac{1}{i \hbar}\sum_{j=3}^{z-2}r[j] \circ r[z+1-j]\right), \;\; z  \geq 4.
\ee 
From   proposition {\bf \ref{tw1}} we see that for the curvature $R_{\Gamma} \neq 0$ there must be $\delta^{-1} R_{\Gamma} \neq 0.$ So on any nonflat Fedosov manifold $({\cal M}, \Gamma)$ the term $r[3]$ is different from $0$.

Assume that $r$ is a finite formal series of the degree $m-1, \; m \geq 4.$ It means that
$r[m-1]$ is the last nonzero term  of the $r$ series. Hence, from (\ref{no1})
\bea
\label{no2}
\delta^{-1}\left( \partial_{\Gamma}r[m-1]+ \frac{1}{i \hbar}\sum_{j=3}^{m-2}r[j] \circ r[m+1-j]\right)& = & 0, \nonumber \\
\delta^{-1}\left(\frac{1}{i \hbar}\sum_{j=3}^{m-1}r[j] \circ r[m+2-j]\right)& = & 0 , \nonumber \\
\vdots & &  \nonumber \\
\delta^{-1}\left(\frac{1}{i \hbar}\big( r[m-2] \circ r[m-1]+ r[m-1] \circ r[m-2]\big) \right) & = &   0,\nonumber \\
\delta^{-1}\left(\frac{1}{i \hbar} r[m-1] \circ r[m-1] \right)& = & 0. 
\eea 
According to theorem  {\bf \ref{no0}} the series $r$ is the only one solution of equation (\ref{f8}). Therefore, relations $\delta r[z]=0, \;\; z \geq m$ imply
\begin{subequations}
\begin{align}
 \partial_{\Gamma}r[m-1]+ \frac{1}{i \hbar}\sum_{j=3}^{m-2}r[j] \circ r[m+1-j]& =  0, \label{no31} \\
\sum_{j=3}^{m-1}r[j] \circ r[m+2-j]& =  0 , \label{no32} \\
\vdots &   \nonumber \\
 r[m-2] \circ r[m-1]+ r[m-1] \circ r[m-2] & =  0, \label{no325}\\
 r[m-1] \circ r[m-1] & =  0. \label{no33}
\end{align}
\end{subequations} 
Conversely, let components $r[z], 3 \leq z \leq m-1,$ where $m \geq 4 $ of the Abelian correction $r$ fulfill
the system of equations (\ref{no31} - \ref{no33}).  Then, applying formula (\ref{no1}) to (\ref{no31}) we see that $r[m]=0.$
Substituting this result and relation (\ref{no32}) to (\ref{no1}) we obtain $r[m+1]=0.$ 
Repeating this procedure we find  that $r[z]=0$ for any $z \geq m.$ Hence $\deg(r) \leq m-1$ so $r$ is the finite formal series.

Concluding,
\begin{tw}
\label{no4}
An Abelian connection $\tilde{\Gamma}= \omega_{ij}X^i dq^j + \Gamma + r$ of the symplectic curvature $2-$form  $R_{\Gamma} \neq 0$ is a finite formal series iff there exists  a natural number $m \geq  4$ such that the components $r[z] , \;\; 3 \leq z \leq m-1,$  of $r$  fulfill the system of equations (\ref{no31} - \ref{no33}). 
\end{tw}
That theorem yields the following statements:
\begin{co}
The sufficient condition for  series $r$ to be infinite is that for any $z$ satisfying inequality $ z \geq 3$
the product $r[z] \circ r[z] \neq 0.$ 
\end{co}
This fact is the straightforward consequence of  equation  (\ref{no33}).
\begin{co}
The sufficient condition for  series $r$ to be infinite is that for any $z$ satisfying inequality $ z \geq 3$ the commutator $\big[r[z],r[z+1]\big]\neq 0.$
\end{co}
The latter conclusion  comes from  (\ref{no325}).
\begin{co}
Let $d \geq  4$ be the minimal value of parameter $m$ for  which the system of equations (\ref{no31} - \ref{no33}) holds.  Then $deg(r)=d-1.$
\end{co}

To illustrate how theorem {\bf \ref{no4}} works, we consider an example of a finite Abelian connection on some symplectic manifold ${\cal M}, \;\; \dim {\cal M} \geq 4$. Detailed anaysis of this case will be presented in the next paper.

\underline{Example}
 
Assume that in some Darboux chart $({\cal U}, \phi)$ on ${\cal M}$   nonzero symplectic connection coefficients \\ $\Gamma_{l_1\,l_2\,l_3}(q^{l_4}, \ldots, q^{l_s}), \;\;  1 \leq l_1, \ldots, l_s \leq \dim {\cal M}$ are these  for which
Poisson brackets $\{q^{l_i},q^{l_j}\}_P=0,$ \\ $ 1 \leq i,j \leq s. $ 
Such connection can be curved if $\dim {\cal M} \geq 4.$  In the considered case all of products $r[z] \circ r[k],$ $ 3\leq z,k $  disappear. Applying (\ref{no1}) we see that 
\[
r[z]= (\delta^{-1}\partial_{\Gamma})^{z-3} \delta^{-1}R_{\Gamma}.
\]
 From theorem {\bf \ref{no4}} for $R_{\Gamma} \neq 0 $  the sufficient and necessary condition for $r$ to be a finite series is that there exists  a natural number $z \geq 4$  satisfying 
\be
\label{ostatek}
(\partial_{\Gamma} \delta^{-1})^{z-3}R_{\Gamma}=0.
\ee
The minimal number $z$ for which  (\ref{ostatek}) holds, is the degree of $r.$


\section{An Abelian connection on a $2$-D phase space}

\setcounter{pr}{0}
\setcounter{co}{0}
\setcounter{tw}{0}
\setcounter{de}{0}
\setcounter{equation}{0}

In this section we prove that any Abelian connection on a curved $2$-D Fedosov space $ ({\cal M}, \Gamma)$ is 
  an infinite formal series.

\begin{pr}
\label{nowy1pr}
Let $({\cal M}, \Gamma)$ be a $2$-D Fedosov manifold and $F \in C^{\infty}({\cal P^*M}[[\hbar]] \otimes \Lambda^2)$ be a $2$-form defined on $({\cal M}, \Gamma)$ fulfilling conditions:
\begin{enumerate}
\item
the form $F$ contains only terms of the same degree so $\exists_{ z \geq 0} \; F=F[z],$ 
\item
the form $F$ contains 
only even powers $\hbar^{2k}, \; k \geq 0$ of the deformation parameter $\hbar.$ 
\end{enumerate}
Then $\delta^{-1}F \circ \delta^{-1}F=0$ iff $F=0.$
\end{pr}
\underline{Proof}
\newline
Computations presented below were also tested in the Mathematica 5.2  by Wolfram Research.  
\newline
`$\Leftarrow $'
\newline
It is obvious, that if $F=0$ then $\delta^{-1}F=0$ and $\delta^{-1}F \circ \delta^{-1}F=0.$
\newline
`$\Rightarrow $'
\newline
 We perform our computations locally in a Darboux chart $(U, \phi)$ and denote canonically conjugated coordinates by $q$ and $p.$ Their Poisson bracket
$
\{q,p\}_{P}=1.
$

In the chart $(U, \phi)$ the most general form $F$ of the degree $(z-1)$ and fulfilling conditions from  proposition {\bf \ref{nowy1pr}} is 
\[
F= \sum_{k=0}^{\left[ \frac{z-1}{4}\right]}\;\sum_{l=0}^{z-1-4k}\hbar^{2k} a_{2k,l \, z-1-4k-l}(q,p)(X^1)^l (X^2)^{z-1-4k-l} dq \wedge dp, \; z \geq 1.
\]
By $a_{2k,l \, z-1-4k-l}(q,p)$ we denote some smooth  functions.
To shorten our notation we omit variables $q$ and $p$.

Hence,
\be
\label{p2}
\delta^{-1}F
=  \sum_{k=0}^{\left[ \frac{z-1}{4}\right]}\;\sum_{l=0}^{z-1-4k}
\frac{\hbar^{2k}}{z-4k+1} \Big( a_{2k,l \, z-1-4k-l}(X^1)^{l+1} (X^2)^{z-1-4k-l}dp -
 a_{2k,l \, z-1-4k-l}(X^1)^l (X^2)^{z-4k-l} dq \Big).
\ee
Let us define
$
b_{2k,l}\stackrel{\rm def.}{=} \frac{1}{z-4k+1}a_{2k,l \, z-1-4k-l }.
$

Therefore
\be
\label{p3}
\delta^{-1}F=  \sum_{k=0}^{\left[ \frac{z-1}{4}\right]}\;\sum_{l=0}^{z-1-4k} \hbar^{2k}
 b_{2k,l }(X^1)^{l+1} (X^2)^{z-1-4k-l}dp -
\sum_{k=0}^{\left[ \frac{z-1}{4}\right]}\;\sum_{l=0}^{z-1-4k} \hbar^{2k}
 b_{2k,l }(X^1)^{l} (X^2)^{z-4k-l}dq. 
\ee

To find the square of (\ref{p3}) we need first to compute the product
\[
(X^{1})^r (X^{2})^j \circ (X^{1})^s (X^{2})^k= \sum_{t=0}^{{\rm min}[r,k]+{\rm min}[j,s]} \frac{1}{t!}\left( \frac{i \hbar}{2}\right)^t \times
\]
\be
\label{a2} 
\times \sum_{a=0}^t (-1)^a \left( \begin{array}{c} t \\ a
\end{array}\right) \frac{r!\: j! \:s!\: k!}{(r-t+a)!\:(j-a)!\:(s-a)!\:(k-t+a)!} (X^{1})^{r+s-t}(X^{2})^{k+j-t}.
\ee
In fact the sum (\ref{a2}) over $a$ different from $0$ can be only elements from  $a={\rm max}[t-r,t-k,0]$ until 
$a={\rm min}[j,s,t].$ Simplifying (\ref{a2}) we see that
\[
(X^1)^r (X^2)^j \circ (X^1)^s (X^2)^k
= r!\: j!\: s!\: k!\: \sum_{t=0}^{{\rm min}[r,k]+{\rm min}[j,s]} \left( \frac{i \hbar}{2}\right)^t (X^1)^{r+s-t}(X^2)^{k+j-t}
\times
\]
\be
\label{a3}
\times \sum_{a={\rm max}[t-r,t-k,0]}^{{\rm min}[j,s,t]}(-1)^a \frac{1}{ a!\: (t-a)!\:(r-t+a)!\:(j-a)!\:(s-a)!\:(k-t+a)!}=
\ee
\be
\label{a9}
 =\sum_{t=0}^{{\rm min}[r,k]+{\rm min}[j,s]}(i \hbar)^t (X^1)^{r+s-t} (X^2)^{k+j-t} f(r,j,s,k,t),
\ee
where
\[
f(r,j,s,k,t) \stackrel{\rm def.}{=} \frac{(-1)^{\rm mx}}{2^t} \frac{r!\:j!\:s!\:k!}{(j-{\rm mx})!(s-{\rm mx})!(t-{\rm mx})!
({\rm mx}-t+r)!({\rm mx}-t+k)!{\rm mx}!} \: \times
\]
\be
\label{a10}
  \times \; _4F_3(\{1,{\rm mx}-j,{\rm mx}-s,{\rm mx}-t\};\{1-t+r+{\rm mx},1-t+k+{\rm mx},1+{\rm mx}\};1).
\ee
In the upper formula  ${\rm mx}\stackrel{\rm def.}{=}{\rm max}[t-r,t-k,0]. $ By $ _4F_3(\{a_1,a_2,a_3,a_4\};\{b_1,b_2,b_3\};x)$ is denoted the generalized hypergeometric function.

Therefore
\[
\delta^{-1}F \circ \delta^{-1}F = \sum_{k=0}^{\left[\frac{z-1}{4}\right]} \;
\sum_{w=0}^{\left[\frac{z-1}{4}\right]}\;
\sum_{l=0}^{z-1-4k}\;
\sum_{r=0}^{z-1-4w}\;
\sum_{u=0}^{\min[l+1,z-4w-r]+\min[z-1-4k-l,r]} i^u \hbar^{2k+2w+u}b_{2k,l}b_{2w,r}\; \times
\]
\[
\times (X^1)^{l+r+1-u}(X^2)^{2z-4k-4w-l-r-u-1} \;\times
\left\{
f(l+1,z-1-4k-l,r,z-4w-r,u) +\right.
\]
\be
\label{n1}
\left.
-f(r,z-4w-r,l+1,z-1-4k-l,u)
 \right\} dq \wedge dp.
\ee
Remember that in $f(l+1,z-1-4k-l,r,z-4w-r,u)$ we put (see (\ref{a10})) ${\rm mx}= {\rm max}[u-l-1,u-z+4w+r,0]$
 and in $f(r,z-4w-r,l+1,z-1-4k-l,u)$ by ${\rm mx}$ we mean ${\rm max}[u-r,u-z+1+4k+l,0].$

From  definition (\ref{a10}) of $f(r,j,s,k,t)$ we see that in the sum (\ref{n1}) only terms with odd $u$ are different from $0$. 
Indeed,  
\[
f(l+1,z-1-4k-l,r,z-4w-r,u)=(-1)^u f(r,z-4w-r,l+1,z-1-4k-l,u).
\]
Hence
\[
\delta^{-1}F \circ \delta^{-1}F=
\sum_{k=0}^{\left[\frac{z-1}{4}\right]} \;
\sum_{w=0}^{\left[\frac{z-1}{4}\right]}\;
\sum_{l=0}^{z-1-4k}\;
\sum_{r=0}^{z-1-4w}\;
\sum_{u=0}^{\left[\frac{\min[l+1,z-4w-r]+\min[z-1-4k-l,r]}{2}-\frac{1}{2}\right]} 
 2i(-1)^u \hbar^{2k+2w+2u+1}b_{2k,l}b_{2w,r}\; \times
\] 
\be
\label{n2}
\times (X^1)^{l+r-2u}(X^2)^{2z-4k-4w-l-r-2u-2} \; 
f(l+1,z-1-4k-l,r,z-4w-r,2u+1) dq \wedge dp.
\ee

The degree of the product 
$
\delta^{-1}F \circ \delta^{-1}F
$
equals $2z$
and each term of (\ref{n2}) is determined by powers of $\hbar^{2A+1}$ and $(X^1)^B$ only. It is obvious that
$
4A+B+2 \leq 2z.
$

Formula (\ref{n2}) can be written as
\be
\label{n3}
\delta^{-1}F \circ \delta^{-1}F=\hbar^{2A+1}(X^1)^B(X^2)^{2z-4A-B-2}g_{2A+1,B},
\ee
where $g_{2A+1,B}$ are some coefficients computed below.

Comparing (\ref{n2}) and (\ref{n3}) we obtain the  system of equations
\be
2k+2w+2u+1=2A+1 \;\;\; , \;\;\; l+r-2u=B.
\ee
Its solutions are
\be
u=A-k-w \;\;\; , \;\;\;r=2 A + B - 2 k  - 2 w-l.
\ee

So
\[
g_{2A+1,B} \sim 
\sum_{k=0}^{\left[\frac{z-1}{4}\right]} \;
\sum_{w=0}^{\left[\frac{z-1}{4}\right]}\;
\sum_{l=0}^{z-1-4k}\;
2i(-1)^{A-k-w}b_{2k,l}b_{2w,2 A + B - 2 k  - 2 w-l}\times
\]
\[
 \times f(l+1,z-1-4k-l,2 A + B - 2 k  - 2 w-l,z-2A-B+2k-2w+l,2A-2k-2w+1).
\]
We did not write the equality symbol because not for all $A,B$ this formula works. The reason is that parametres 
$u$ and $ r$
fulfill conditions
\bea
\label{n4}
&0 \leq 2u \leq \min[l+1,z-2A-B+2k-2w+l]+\min[z-1-4k-l,2 A + B - 2 k  - 2 w-l]-1,& 
\nonumber \\ 
& & \nonumber \\
   & 0  \leq r \leq z-1-4w.& 
\eea
which were not taken into account in sums appearing in the definition of $g_{AB}.$
From inequalities (\ref{n4}) we obtain that
\be
\label{n5}
k+w \leq A\;\;,\;\; |k-w| \leq \left[ \frac{z-1}{2}\right]-A
\ee
and
\be
\label{n5.5}
2A+B-z+1-2k+2w \leq l \leq 2A+B-2k-2w.
\ee

Finally 
\[
g_{2A+1,B} = \sum_{k=0}^{\min \left[A,\left[ \frac{z-1}{4}\right]\right]} \;\;\sum_{w=\max \left[0,A+k-\left[ \frac{z-1}{2}\right]\right]}^{\min \left[\left[ \frac{z-1}{4}\right],A-k,k-A+\left[ \frac{z-1}{2}\right] \right]} \;\;\sum_{l=\max[0,2A+B-z-2k+2w+1]}^{\min[z-1-4k,2A+B-2k-2w]} 2i(-1)^{A-k-w}\times
\]
\be
\label{n6}
 \times  b_{2k,l}b_{2w,2A+B-2k-2w-l} f(l+1,z-1-4k-l,2 A + B - 2 k  - 2 w-l,z-2A-B+2k-2w+l,2A-2k-2w+1).
\ee

We look for solutions $\delta^{-1}F$ of the equation $\delta^{-1}F \circ \delta^{-1}F=0.$ This equation is equivalent to 
the   system of equations   
\be
\label{nowy6}
g_{2A+1,B}=0, \;\; \;\;4A+B+2 \leq 2z.
\ee
 for functions  $b_{2k,r}.$

We start solving (\ref{nowy6}) from the equation parametrized by
$A=B=0.$
 Since the only one possibility is $k=w=l=0,$ from (\ref{n6}) we immediately obtain 
\[
2i \:b_{0,0}^2\:f(1,z-1,0,z,1)=0.
\]
As 
$
f(1,z-1,0,z,1)= \frac{z}{2}  \neq 0 
$
  we deduce that $b_{0,0}=0.$

Next equation we choose to solve is for $A=0, B=1.$ It contains  only a product $b_{0,0} b_{0,1}$ with some factor  so, due to the fact that $b_{0,0}=0,$ it is fulfilled automatically. The next equation determined by $A=0, B=2$ with the condition $b_{0,0}=0$  reduces to 
\[
g_{1,2}=2i \:b_{0,1}^2\:f(2,z-2,1,z-1,1)=0.
\]
Because 
$
f(2,z-2,1,z-1,1)=\frac{z}{2} \neq 0
$
we conclude that $b_{0,1}=0.$

Repeating that procedure for $A=0, \;B \leq 2z-2$ we see that all 
\be
\label{o3}
b_{0,l}=0\;\;\;,\;\; \; 0 \leq l \leq z-1.
\ee 
Putting $A=1$ we do not receive  conditions to find any other coefficients. But for $A=2, \; B=0$ (that implies $z \geq 5 $) we see that if relations (\ref{o3}) hold then
\[
g_{5,0}=2i \:b_{2,0}^2\:f(2,z-5,0,z-4,1)=0.
\]
As
$
f(2,z-5,0,z-4,1)= z-4 \neq 0,
$
we see that $b_{2,0}=0.$ 
Following this way we solve the system of equations (\ref{nowy6}) completely obtaining all of $b_{2k,r}=0.$  

Thus we showed that if $\delta^{-1}F \circ \delta^{-1}F=0$ then $F=0.$
 \hspace{1cm} \rule{2mm}{2mm}

\vspace{0.5cm}
As it has been said in  subsection \ref{fac},  on a $2$-D Fedosov  manifold $({\cal M}, \Gamma)$ relation $R_{\Gamma} \neq 0$ yields $r[3] \neq 0.$
Hence $r[3]= \delta^{-1}R_{\Gamma},$
from proposition {\bf \ref{nowy1pr}} we obtain that $r[3] \circ r[3]\neq 0.$ Using theorem {\bf \ref{no4}} we conclude that there exists at least one nonzero component $r[z]$ of the correction $r$ of the degree $z > 3.$ Remembering that (see (\ref{no1}))
$r[z]= \delta^{-1}(F[z-1]),$ where $F[z-1]\stackrel{\rm def.}{=}\partial_{\Gamma}r[z-1]+ \frac{1}{i \hbar}\sum_{j=3}^{z-2}r[j] \circ r[z+1-j]$
and then
applying proposition {\bf \ref{nowy1pr}} to  $r[z]$ we see that $r[z] \circ r[z]\neq 0.$ Hence, from  theorem {\bf \ref{no4}} there exits   $r[z_1] \neq 0$ such that $z_1 > z.$ Definition (\ref{no1}) of $r[z_1]$ plus proposition {\bf \ref{nowy1pr}} guarantiees that $r[z_1] \circ r[z_1]\neq 0.$ Hence, from theorem {\bf \ref{no4}}   there must be $\deg(r)>z_1.$ Following this pattern  
 we arrive at the following
\begin{tw}
\label{p6}
On $2$-D phase space with nonvanishing symplectic curvature $2$-form $R_{\Gamma}$ any Abelian connection is an infinite series.
\end{tw}

We stress that theorem {\bf \ref{p6}} holds for $2$-D real symplectic manifolds
with the correction $r$ determined by formula (\ref{2}) fulfilling (\ref{h1}). On  K\"{a}hler locally symmetric manifolds with the complex dimension $1$ there exists a finite Abelian connection (see \cite{tama}) but for it $\delta^{-1}r \neq 0.$ 

\section{Conclusions}

Fedosov quantization method is based on recurrent formulas (\ref{2}) and (\ref{nowy09}). The first one  defines the correction $r$ to the Abelian connection, the second   flat section $a \in {\cal P^*M}[[\hbar]]_{\tilde{\Gamma}}   $ representing quantum observable $a_0$. 
There is no general rule saying in which cases the flat section $a$ or the $*$-product of functions $a_0 * b_0$ can be written in a compact form. Such situation happens inter alia
when  both  iterations (\ref{2}) and (\ref{nowy09}) generate finite formal series. 

In current paper we consider the question, when the Abelian connection on a Fedosov manifold $({\cal M}, \Gamma)$ described in  theorem {\bf \ref{no0}} is a finite formal series. We find a system of equations fixing the sufficient and necessary condition for $r$ to be finite. 

Then we  apply the result quoted above to the case 
  of $2$-D phase space with nonvanishing curvature.
We prove that 
 the series $r$ on such spaces is always infinite. By the way  we find an explicit formula describing the $\circ$-product in $2$-D case.

\end{document}